\documentclass[aps,superscriptaddress,showpacs]{revtex4}
\usepackage{amsmath}
\usepackage{graphicx}
\begin{document}
\title{Nonlinear dynamics of large amplitude dust acoustic shocks and solitary pulses in dusty plasmas}
\author{P. K. Shukla}
\affiliation{International Centre for Advanced Studies in Physical Sciences \& Institute for Theoretical Physics, Faculty of Physics \& Astronomy, Ruhr University Bochum, D-44780 Bochum, Germany}
\affiliation{ Department of Mechanical and Aerospace Engineering \& Centre for Energy Research, University of California San Diego, La Jolla, CA 92093, U. S. A.}
\email{profshukla@yahoo.de}
\author{B. Eliasson}
\affiliation{International Centre for Advanced Studies in Physical Sciences \& Institute for Theoretical Physics, Faculty of Physics \& Astronomy, Ruhr University Bochum, D-44780 Bochum, Germany}
\email{bengt@tp4.rub.de}
\received{25 May 2012}
\revised{19 September 2012}
\begin{abstract}
We present a fully nonlinear theory for dust acoustic (DA) shocks and DA solitary pulses
in a strongly coupled dusty plasma, which have been recently observed experimentally by
Heinrich {\it et al.} [Phys. Rev. Lett. {\bf 103}, 115002 (2009)],  Teng {\it et al.}
[Phys. Rev. Lett. {\bf 103}, 245005 (2009)], and Bandyopadhyay {\it et al.}
[Phys. Rev. Lett. {\bf 101}, 065006 (2008)]. For this purpose, we use a generalized
hydrodynamic model for the strongly coupled dust grains, accounting for arbitrary large
amplitude dust number density compressions and potential distributions associated with
fully nonlinear nonstationary DA waves. Time-dependent numerical solutions of
our nonlinear model compare favorably well with the recent  experimental works
(mentioned above) that  have reported the formation of large amplitude non-stationary
DA shocks and DA solitary pulses in low-temperature dusty plasma discharges.
\end{abstract}

\maketitle

\section{Introduction}

Charged dust grains and dusty plasmas \cite{Goertz89,Shuklamamun,Mendis02,Vladimirov05,Fortov05,Shukla09,Fortov10} are ubiquitous
in astrophysical environments (e.g. interstellar media, molecular dusty clouds, star forming clouds, supernovae such as the Eagle Nebula, etc.),  in planetary ring systems \cite{Goertz89,Horanyi04}  (e.g. the spokes in Saturn's rings recorded by  the Voyager spacecraft cameras), in our solar system (e.g. interplanetary dust particles produced by comets), as well as  near the Sun's and Earth's atmospheres  (e.g. the mesospheric and ionospheric regions). Charged dust particles are naturally formed in industrial processing of nanotechnology and in magnetic fusion reactors.

It is well-known that charging of a neutral dust particle occurs due to a variety  of physical processes \cite{Merlino94,Mendis94},  including the collection of electrons from the background plasma, photo emissions, tribo-electric effects, etc. In the remote past, it was shown by Wuerker {\it et al.} \cite{Wuerker} that an ensemble of electrically charged iron and aluminum particles having diameters of a few microns
can be confined  by three-dimensional focusing forces of alternating and static electric fields and the Coulomb repulsion, leading eventually to the formation of crystallized arrays of ions and aluminum dust particles, which can be melted and reformed. However, a dusty plasma is usually composed of electrons, positive ions, negative or positive dust grains, and neutral atoms. When  the interaction potential energy ($=Z_d^2 e^2/d$, where $Z_d$ is the dust charge state, $e$ the magnitude of  the electron charge, and $d$ the inter-dust grain  distance or the Wigner-Seitz radius) between two neighboring dust  grains is much larger (smaller) than the dust kinetic energy  $k_B T_d$, where $k_B$ is the Boltzmann constant and $T_d$ the dust temperature, the dusty plasma is in a strongly (weakly) coupled state.  Following the charged particles condensation idea \cite{Ichimaru82} of one component strongly correlated electron system, Ikezi \cite{Ikezi86} postulated the solidification of charged dust particles when the dusty plasma $\Gamma=Z_d^2e^2 \exp(-d/\lambda_D)/d k_B T_d$ exceeds 172,  taking into account the plasma screening effect, where $\lambda_D$  is the plasma Debye radius \cite{Shuklamamun}. Such values of $\Gamma$ can be achieved in low-temperature laboratory discharges at room temperatures owing to the large $Z_d$ acquired by a micron-size dust grain by  absorbing electrons from the background plasma.
There are also Monte-Carlo and Molecular Dynamics simulations that accurately depict different states
of ordered  dust  structures \cite{Hamaguchi97,Nosenko09,Ivlev08} when dust grains are repelling each other according to the Yukawa or Debye-H\"uckel  force. The phase diagram for  $\Gamma_s$ against
$\kappa$ = average inter-dust grain  spacing/dusty plasma Debye radius,  indeed reveal dust solid face-centered cubic  (fcc), dust solid  body-centered cubic (bcc) and  dust fluid phases for a set of  $\Gamma_s$ and $\kappa$ values, as given in Ivlev {\it et al.} \cite{Ivlev08} where an empirical scaling for dust crystal melting is also given.

The formation  of dust Coulomb crystals and ordered dust structures have been observed in the sheath region of many laboratory experiments \cite{Chu1,Chu2,Thomas,Hayashi},
where charged dust grains are kept together due to confining electrostatic potentials in a plasma sheath.
However, robust ordered dust structures may also be formed due to attractive forces \cite{Shuklamamun} between negative  dust grains associated with ion focusing and ion wakefields \cite{Nambu95,Shukla96}
in a dusty plasma sheath with streaming ions, shadowing forces due to collisions with ions \cite{Ignatov96,Lampe00}, as well as due to  overlapping Debye spheres \cite{Resendes98} and dipole-dipole interactions \cite{Lee97,Mohideen98,Tskhakaya04}.  The alignment of charged dust grains in an assembly due to the attractive force associated with ion focusing and ion wakefield effects has been experimentally  observed \cite{Takahashi}. Furthermore, the collective behavior of dusty plasmas involving an ensembles of charged dust grains was recognized through the  prediction of  the dust acoustic wave (DAW) by Shukla \cite{Capri}
at the First Capri Workshop  on Dusty  plasmas in May of 1989, where he suggested the existence of the nonlinear DAW in the presence of  Boltzmann  distributed electrons and ions,  and massive, charged dust particles. This idea was then worked out  in the  first paper \cite{Rao} on the DAW. It must be  stressed that there does not exist a counterpart of the DAW in an electron-ion plasma without charged dust  grains, since the DAW is supported by the dust particle inertia, and the restoring force comes from the  pressures of the inertialess hot electron and ions. Thus, similar to the Alfv\'en wave in a magnetized plasma,  the DAW is of fundamental importance in laboratory and space plasmas physics. The DAW is usually excited by an ion streaming instability, and has a frequency much smaller than the dusty plasma frequency, extending into the infra-sonic frequency range. Low-frequency (of the order of 10 Hz) DA  fluctuations were first observed in the experiment  of  Chu  {\it et al.} \cite{Chu1}, and have since been observed in many laboratory experiments world-wide \cite{Shuklamamun,Shukla09,Chu1,Barkan,Fortov,Prab,Chang12}, and also in the Earth's ionosphere \cite{Popel09}.

Ichimaru {\it et al.} \cite{Ichimaru} further extended the theory of strong coupling and viscosity coefficients for a high-density one component electron plasma.  Berkovsky \cite{Berk} developed a generalized hydrodynamic model for plasmas with strongly coupled ions and degenerate electrons, and used it to
investigate the linear properties of modified ion-acoustic waves. A similar theory was developed for strongly
correlated dust grains in dusty plasmas by Kaw and Sen \cite{Kaw}, who presented a generalized
viscoelastic hydrodynamic model for strongly correlated dust grains  and investigated the linear properties of dust acoustic waves,  especially the low-frequency longitudinal and transverse modes in a strongly coupled dusty plasma.
The latter model has also been extended to the  weakly nonlinear regime \cite{Veeresha10} to study the propagation of small amplitude nonlinear dust acoustic waves in a strongly coupled dusty plasma.

However, recently a number of laboratory experiments \cite{Fortov05b,Heinrich,Merlino,Band,Teng} have reported observations of nonlinear DAWs in the form of extremely large amplitude DA shocks \cite{Fortov05b,Heinrich,Merlino} and DA solitary pulses \cite{Band,Teng,Chang12} at  kinetic levels. Physically, the large amplitude DA shocks are formed when nonlinearities in plasmas balance the DAW dissipation caused  by the dust fluid viscosity coming from dust grain correlations in strongly coupled dusty plasmas, while DA solitary pulses arise in  the collisionless regime due to the balance between the harmonic generation nonlinearities and the DAW dispersion. To the best of our knowledge, there are no theories for arbitrary large amplitude nonlinear, nonstationary DA shocks and DA solitary pulses in dusty plasmas with dust correlations. It should be stressed that small amplitude theories for DA shocks and DA solitary pulses based on the Burgers \cite{Mamun09}, Korteweg-de Vries (KdV), and KdV-Burgers equations \cite{Veeresha10} are not suitable for explaining observations \cite{Fortov,Band,Heinrich,Teng, Merlino} that report anomalously high (up to $40\%$  and beyond) dust density compressions.  A large amplitude theory of Eliasson and Shukla \cite{Eliasson} for
a collisionless dusty plasma explains well the DAW steepening and nonlinear wave speed \cite{Heinrich,Merlino}, but is unable to predict the shock width observed in the experiments.

In this paper, we present a fully nonlinear, non-stationary unified theory for arbitrary large amplitude DA shocks and DA solitary pulses in a dusty plasma, taking into account the effects of strong coupling between charged dust  grains, the nonlinear polarization force acting on charged dust grains due to thermal ions that shield negative dust grains, collisions between charged dust grains and neutrals,  dust correlations decay rate, the dust fluid shear and bulk viscosities, etc. This gives a more complete picture of various non-ideal effects in dusty plasmas, and we are thus able to provide a comparison between our new non-stationary and fully nonlinear theory with the recent laboratory observations of DA shocks and DA solitary pulses \cite{Band,Heinrich,Teng,Merlino}. Neglected are effects due to attractive forces (ion focusing, wake fields, etc.) between dust grains,  which may affect the equation of state and transport coefficients of the system. These effects, however, are either small or depend on the moment transfer of streaming ions,
which we do not consider here.

\section{Mathematical model}

We consider a dusty plasma composed of inertialess electrons and ions, as well as strongly correlated negatively charged micron-sized dust particles of uniform sizes. In the presence of large amplitude ultra-low frequency DA waves, with $ \omega \ll \nu_{en}, \nu_{in} \ll  k^2 V_{Te, Ti}^2/\omega$, where $\omega$ is the wave frequency,  $\nu_{en}$ $(\nu_{in})$ the electron (ion)-neutral collision frequency,   $k$ the wave number, and $V_{Te}$ $(V_{Ti})$ the electron  (ion) thermal speed. Both electrons and ions follow the Boltzmann law, since they  can be considered inertialess on the timescale of the DAW period, and henceforth rapidly thermalize under the action of collisions. Thus, the electron and ion number densities are, respectively,  $n_e =n_{e0} \exp(e\phi/k_B T_e)$, and $n_i =n_{i0} \exp(-e\phi/k_B T_i),$ where
$n_{e0}$ and $n_{i0}$  are the unperturbed electron and ion number densities, respectively, $e$ the magnitude of the electron charge,  $\phi$ the electrostatic potential, $k_B$ the Boltzmann constant, and $T_e$ $(T_i)$ the electron  (ion) temperature. At equilibrium, we have  the quasi-neutrality condition $n_{i0} =n_{e0} + Z_d n_{d0}$, where  $Z_d$ is the average number of electrons residing on a dust grain,
and $n_{d0}$ the unperturbed  dust number density.

The dust particle dynamics associated with fully nonlinear, non-stationary DAWs in a strongly coupled dusty plasma is governed by the   generalized hydrodynamic equations composed of the  dust continuity equation $(\partial n_d/\partial t)+ \nabla \cdot (n_d {\bf v}_d) =0$,  and the generalized dust momentum equation
\begin{eqnarray}
 \left(1+ \tau_r \frac{d}{dt} \right) \left[\frac{d{\bf v}_d} {dt} + \nu_d {\bf v}_d
- \frac{Z_d e}{m_d}\nabla \phi + \frac{Z_d e R}{m_d} \left(\frac{n_i}{n_{i0}}\right)^{1/2} \nabla \phi
 + \frac{k_B T_d}{\rho_d}\nabla\bigg(\mu_d n_d \bigg) \right]\nonumber \\
= \frac{\eta}{\rho_d} \nabla^2 {\bf v}_d + \frac{\left(\xi + \frac{\eta}{3}\right)}{\rho_d}
 \nabla (\nabla \cdot {\bf v}_d),
\end{eqnarray}
taking into account finite amplitude convective and pressure nonlinearities \cite{Eliasson}, nonlinear ion polarization force,  strong dust coupling effects \cite{Ichimaru,Berk,Kaw,Frenkel}, and dust neutral collisions \cite{Baines65}. Here $d/dt =(\partial t/\partial t) + {\bf v}_d \cdot \nabla$ is the total time derivative, $n_d$ and  ${\bf v}_d$ are the dust number density and dust fluid velocity, respectively, $m_d$ the dust mass,  $\rho_d =n_d m_d$ the dust mass density, $R =Z_de^2/4 k_B T_i \lambda_{Di}$ is a parameter determining the effect of the polarization force \cite{Khrapak}, which reduces the phase speed of the DAW,  arising from interactions between thermal ions and negative dust grains,
$\mu_d n_d k_B T_d\equiv P_d$ the effective dust thermal pressure for a one component plasma (OCP) \cite{Kaw}, where $\mu_d =1+(1/3)u(\Gamma) +(\Gamma/9)\partial u(\Gamma)/\partial \Gamma$  the compressibility, $\Gamma=Z_d^2 e^2/d k_B T_d$ the ratio between the dust Coulomb and dust thermal energies,  $d = (3/4\pi n_{d0})^{1/3}$ the Wigner-Seitz dust grain separation distance, and $u(\Gamma)$ is a measure of the excess internal energy of the system, which reads \cite{Abe,Slatt} $u(\Gamma)\simeq -(\sqrt{3}/2)\Gamma^{3/2}$ for $\Gamma \leq 1$ ({\it viz}. a liquid-like state), and $u(\Gamma) =-0.80 \Gamma + 0.95 \Gamma^{1/4} + 0.19 \Gamma^{-1/4} -0.81$ in a range $1 < \Gamma < 200$. Furthermore,  the effective dusty plasma Debye radius $\lambda_D= \lambda_{De}\lambda_{Di}/(\lambda_{De}^2+\lambda_{Di}^2)^{1/2}$, where $\lambda_{De}=(k_B T_e/4 \pi n_{e0} e^2)^{1/2}$ and $\lambda_{Di}=(k_B T_i/4 \pi n_{i0} e^2)^{1/2}$ are the ion and electron Debye radii, respectively. The dust-neutral collision frequency is given by the Epstein formula \cite{Baines65}  $\nu_{dn}=(8/3)\sqrt{2\pi} m_n n_n r_d^2 v_{Tn}/m_d$, where  $m_n$ is the neutral mass, $n_n$ the neutral number density, $r_d$ the dust grain radius, $V_{Tn}=(k_B T_n/m_n)^{1/2}$ the neutral  thermal speed, and $T_n$ the neutral gas temperature.  The visco-elastic properties of the dust fluids are characterized by the dust correlation relaxation time \cite{Ichimaru,Berk}  $\tau_r =[ (\xi +4\eta/3)/n_{d0}T_d]/\left[1-\mu_d
+ 4 u(\Gamma)/15\right]$, involving the shear and bulk viscosities $\eta$ and $\xi$, respectively. There are various approaches for calculating $\eta$ and $\xi$, which are widely discussed in the literature \cite{Slatt}.
The DA wave potential $\phi$ is obtained from Poisson's equation $\nabla^2 \phi =4 \pi e (n_e -n_i + Z_d n_d)$, taking into account the dispersive effect due to the departure from quasi-neutrality.
The ion drag force \cite{Barnes92,Khrapak05,Hutchinson06}  acting on a dust grain has been neglected in  Eq. (1), which is justified  since the ions are assumed to follow the Boltzmann distribution with no ion momentum flow.  However, in a dusty plasmas with large dust particles and equilibrium ion flows,
there can be an instability with a growth rate much smaller than the DAW frequency \cite{Shuklamamun}.

\section{One-dimensional quasi-stationary shocks and solitary waves}

Let us now consider the simplest problem of one-dimensional nonlinear DAWs propagating along the $x$-axis in a Cartesian coordinate system. We define the dimensionless variables $N= n_d/n_{d0}$, $U = \hat {\bf x}  \cdot {\bf v}_d/C_d$, and $\Phi = e \phi/k_B T_i$,  where $C_d=\omega_{pd}\lambda_D$ is the
dust acoustic speed, $\omega_{pd}=(4\pi n_{d0} Z_d^2 e^2 /m_{d0})^{1/2}$ the dust plasma frequency,
and $\hat {\bf x}$ the unit vector along the $x-$axis.  We then have the dust continuity equation
\begin{equation}
\frac{DN}{DT} + N \frac{\partial U}{\partial X} =0,
\end{equation}
the generalized viscoelastic dust momentum equation
\begin{equation}
\left(1+ a \frac{D}{DT}\right)\left[\frac{DU}{DT} + \nu U - \left[1- R\exp(-\Phi/2)\right] \frac{\gamma}{P} \frac{\partial \Phi}{\partial X}
+ T_0 \frac{\partial {\rm ln }N}{\partial X}\right]- \frac{\beta}{\Lambda} \frac{\partial^2 U}{\partial X^2} =0,
\end{equation}
and Poisson's equation
\begin{equation}
\gamma \frac{\partial^2 \Phi}{\partial X^2} = (1-P) \exp(\tau \Phi)- \exp(-\Phi) + P N,
\end{equation}
where $a =\omega_{pd}\tau_r$, $\nu =\nu_{dn}/\omega_{pd}$, $D/DT =\partial/\partial T + U \partial/\partial X$, $T=\omega_{pd}t$,  $X =x/\lambda_D$, $\Lambda=\lambda_D^2/d^2$,
$\beta =(\xi + 4\eta/3)/m_d n_{d0} \omega_{pd} d^2$ (typical values \cite{Ichimaru}   of $\beta$ are  roughly $1.04$, $0.08$, and $0.3$ for $\Gamma =1$, $10$ and $160$, respectively), $T_0= \mu_d T_d \gamma/Z_d T_i P$,  $\gamma=1+\tau(1-P)$, $P=Z_d n_{d0}/n_{i0}$, and $\tau=T_i/T_e$. We are
assuming here that the constant parameter $P$ is given for a set of experiments; however, it has been experimentally shown \cite{Merlino94} that $Z_d$ is typically reduced for closely packed   ($d < \lambda_D$) dust grains. This effect, which can be important at high dust number densities, will be neglected here for simplicity.  Furthermore, the dust charge fluctuation effect has been neglected, since the dust charging time-period ($\nu_1^{-1}$) is usually much shorter  than the  time period for the formation of nonlinear DAWs
 we are concerned with \cite{Shukla12}, and the  fugacity parameter ${\cal F}  = 4\pi n_{d0}\lambda_{Di}^2 r_d\nu_2/\nu_1 (1+ n_{e0}T_i/n_{i0} T_e) $ is smaller than 1,  where the expressions for $\nu_1$ and $\nu_2$ are given  in Refs. \cite{Shuklamamun,Shukla12}.

In a stationary frame such that all physical variables depend only on $\zeta=X-MT$ with $M =U/C_d$, where $U$ is the constant speed of the nonlinear DA waves, we have $U =M(N-1)/N$, so that the dust momentum equation (3) reads
\begin{equation}
\begin{split}
&\bigg(1-\frac{a M}{N}\frac{\partial}{\partial\zeta}\bigg)\bigg[\frac{M^2}{2}\frac{\partial}{\partial\zeta}\bigg(\frac{1}{N^2}\bigg)
+\nu M\frac{(N-1)}{N} - [1-R\exp(-\Phi/2)] \frac{\gamma}{P} \frac{\partial\Phi}{\partial \zeta}
+ T_0 \frac{\partial {\rm ln} N}{\partial \zeta}\bigg] \\
&+\frac{\beta M}{\Lambda} \frac{\partial^2}{\partial \zeta^2}\bigg(\frac{1}{N}\bigg) =0,
\end{split}
\end{equation}
which couples with Poisson's equation
\begin{equation}
\gamma\frac{\partial^2 \Phi}{\partial \zeta^2} = (1-P) \exp(\tau \Phi)-\exp(-\Phi) + P N.
\end{equation}

Quasistationary DA shock waves exist only for $\nu=0$, when the dust-neutral collisions can be neglected. Furthermore, it is possible to derive  a simple condition for the DA shock wave amplitudes depending on other parameters when  the relaxation time for dust grain correlations is  much smaller than the dust plasma period. Hence, for $a=\nu=0$, Eq. (5) can be integrated once to obtain
\begin{equation}
\frac{M^2}{2}\bigg(\frac{1}{N^2}-1\bigg)-\frac{\gamma}{P} \Phi + \frac{2 \gamma R}{P}[1-\exp(-\Phi/2)]
+  T_0{\rm ln} N+\frac{\beta M}{\Lambda}\frac{\partial}{\partial \zeta}\bigg(\frac{1}{N}\bigg)=0,
\end{equation}
where we have used the boundary conditions $N=1$, $\Phi=0$ and $\partial/\partial \xi=0$ at $\zeta=+\infty$.  The DA shock amplitude  at $\zeta=-\infty$, where $\partial/\partial \xi=0$, $N=N_{shock}>1$ and $\Phi=\Phi_{shock}<0$ is now obtained from Eq. (7) as
\begin{equation}
\frac{M^2}{2}\bigg(\frac{1}{N_{shock}^2}-1\bigg)-\frac{\gamma}{P}\Phi_{shock}
+\frac{2\gamma R}{P}[1-\exp(-\Phi_{shock}/2)]+T_0{\rm ln} N_{shock} =0,
\end{equation}
while Eq. (6) yields
\begin{equation}
N_{shock}=\frac{\exp(-\Phi_{shock})-(1-P)\exp(\tau \Phi_{shock})}{P}.
\end{equation}
Using Eq. (9) we can eliminate $N_{shock}$ from Eq. (8) to obtain $M$ as a function of the shock wave potential  $\Phi_{shock}$ for the parameters $R$, $T_0$, $P$, and $\tau$.  The term proportional to $\beta/\Lambda$ in Eq. (7) works  to smoothen the shock front, but does not influence the shock amplitude.  The DA shocks are associated with a positive  jump of the dust number density, $N_{shock}>1$, and  a decrease of the potential, $\Phi_{shock}<0$, for $M > C_a$, where  $C_a=(1-R+T_0)^{1/2}$ is the linear DAW speed in the long-wave limit $\partial/\partial \zeta=0$. Hence, the DA shocks are propagating with super-DA speeds in comparison with the upstream plasma.

\begin{figure}[htb]
\centering
\includegraphics[width=8.5cm]{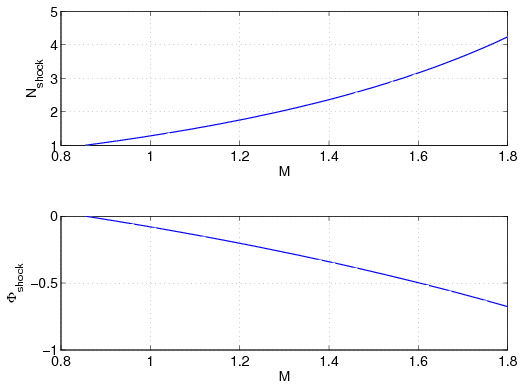}
\caption{The DA shock potential and associated dust number density as a function of $M$ for $P =0.3$,
$\tau= 0.012$, $R=0.28$, and $T_0=0.01$. The DA shock potential is negative for increasing dust
number density. The amplitudes increase with the increase of $M$.}
\label{Fig:amplitude}
\end{figure}

\begin{figure}[htb]
\centering
\includegraphics[width=12cm]{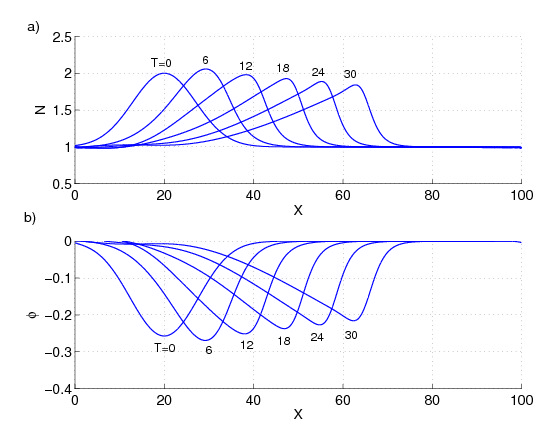}
\caption{The time and space evolution of (a) the dust number density and (b) the DA wave potential for $a=0.01$, $\beta=0.15$, $\Lambda=0.18$, $\nu=0.002$, $P=0.3$, $R=0.28$, $T_0=0.01$, and $\tau=0.012$, corresponding to the plasma parameters of Refs. \cite{Heinrich,Merlino}.}
\label{Fig:Merlino}
\end{figure}

\section{Comparison with experiments}

In Figs. \ref{Fig:amplitude} and \ref{Fig:Merlino}, we have used the plasma parameters of Refs. \cite{Heinrich,Merlino} to study the nonlinear dynamics and the formation of shocks involving large
amplitude  DA pulses. The parameters of the experiment \cite{Heinrich} are $n_i=2\times 10^{14}\,\mathrm{m}^{-3}$, $T_i=0.03\,\mathrm{eV}$, $T_e=2.5\,\mathrm{eV}$,  $Z_d=2\times10^3$, $n_d=3\times10^{10}\,\mathrm{m}^{-3}$, $m_d=10^{-15}\,\mathrm{kg}$, $r_d=0.5\,\mathrm{\mu m}$, giving $\omega_{pd}=590\,\mathrm{s}^{-1}$, $\lambda_D\approx 85\mathrm{\mu m}$, $C_d=50\,\mathrm{mm/s}$, and $d\approx 2\times 10^{-4}\,\mathrm{m}$. The used gas (argon, $m_n=3.6\times10^{-29}\,\mathrm{kg}$) at the pressure 13 Pa and temperature $T_n=0.03\,\mathrm{eV}$ gives a neutral number density $n_n=3\times 10^{21}\,\mathrm{m}^{-3}$, and  a dust-neutral collision frequency $\nu_{dn}\approx 1\,\mathrm{s}^{-1}$.
It was observed in the experiment \cite{Heinrich} that a large amplitude dust density pulse self-steepened and formed a shock-like structure, which propagated with a mean speed of about $75\,\mathrm{mm/s}$, somewhat higher than the estimated dust acoustic speed.
 For the given parameters, we have $\Lambda=0.18$, $R=0.28$, $P=0.3$, and $\tau=0.012$. The normalized dust-neutral collision frequency  $\nu\approx 3\times10^{-3}$
is quite small, while the dust fluid viscosity due to strong dust coupling effects is more prominent.
We choose $\beta=0.15$, which is compatible with the experimental $\Gamma \gtrsim 1$. In addition, we choose $a=T_0=0.01$.  Figure \ref{Fig:amplitude} displays $M$ as a function of dust number density and associated potential, obtained from Eqs. (8) and (9).  In the small amplitude limit, viz. $N_{shock}\rightarrow 1$  and $\Phi_{shock}\rightarrow 0$, we have $M\rightarrow C_a\approx 0.85$. The DA shock speed $M$ increases with increasing DA shock wave amplitudes, with an increase of the dust density and an associated negative potential. Figure \ref{Fig:Merlino} shows a simulation of the time-dependent system of Eqs. (5)--(8). As initial conditions, we used $N=1+\exp[-(X-20)^2/100]$ and $U=0.7\exp[-(X-20)^2/100]$. The profiles of the dust number density and DAW potential in Fig. \ref{Fig:Merlino} show that the initial DA pulse steepens and a monotonic DA shock is formed, similar to the one in Fig. 5 of Ref. \cite{Heinrich}. The large amplitude (100\%) dust density  perturbations are associated with a negative potential $\Phi\approx -0.25$.  The average speed of the DA density pulse is $M\approx 1.4$, in good agreement with Fig. 1 for $N_{shock}\approx 2$, $M\approx 1.3$, and $\phi_{shock}\approx -0.25$. We found  that monotonic (oscillatory) DA shocks exist for $\beta\gtrsim \Lambda$ ($\beta\lesssim \Lambda$),  and solitary waves in the limit $\beta\ll \Lambda$. In dimensional units, the simulated nonlinear wave speed is about $70\,\mathrm{mm/s}$, which is close to the experimental value in Ref. \cite{Heinrich}.

\begin{figure}[htb]
\centering
\includegraphics[width=12cm]{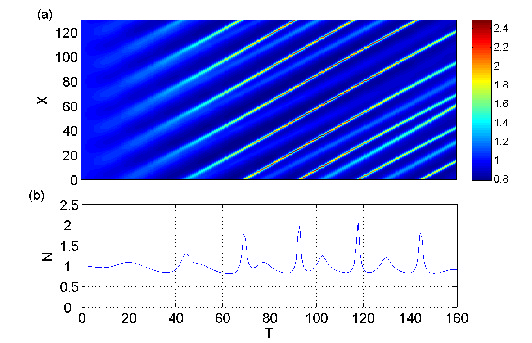}
\caption{a) The time and space evolution of the dust number density for  $a=\beta=\nu=R=T_0=0$, $P=0.51$, and $\tau=0.025$.   b) The time variation of $N$ at $X=0$. The driven DAW develops into spiky solitary DAW structures similar to those  observed by  Teng {\it et al.} \cite{Teng}.}
\label{Fig:Teng}
\end{figure}

\begin{figure}[htb]
\centering
\includegraphics[width=12cm]{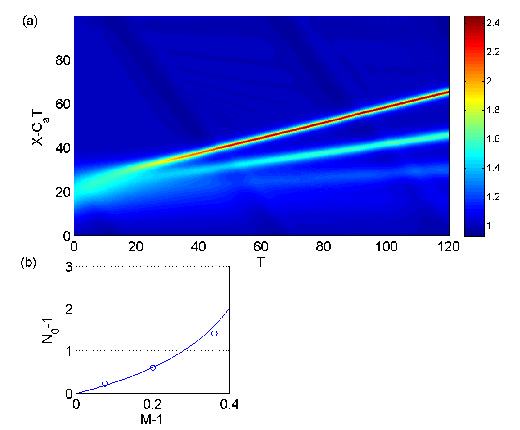}
\caption{a) The time and space evolution of the dust number density $N$ for  $a=\beta=\nu=R=T_0=0$, $P=0.43$, and $\tau=0.038$.   The initial broad pulse breaks up into three separate DA solitary pulses propagating with the super-acoustic speed, similar to those observed  by Bandyopadhyay {\it et al.} \cite{Band}. b) A comparison between the soliton amplitude obtained numerically (circles) with the
theoretical  amplitude $N_0$ (solid line).}
\label{Fig:Band}
\end{figure}

We next turn to laboratory observations of large-amplitude localized DA solitary pulses in weakly
collisional plasma discharges. Teng {\it et al.} \cite{Teng} and Chang {\it et al.} \cite{Chang12}
observed the formation of large amplitude, localized dust density structures, driven by  a flow of
ions towards the bottom of the plasma discharge.
From the given parameters \cite{Teng,Chang12}
$n_e=10^9\,\mathrm{cm}^{-3}$, $T_e=4\,\mathrm{eV}$,
$n_i=1.2\times10^9\,\mathrm{cm}^{-3}$, $T_i=0.05\,\mathrm{eV}$,
$Z_d=5000$, $n_{d}\approx 3.7\times10^4\mathrm{cm}^{-3}$ (inter-dust distance about $0.3\,\mathrm{mm}$),
and $m_d=6.9\times 10^{-11}\mathrm{g}$, we have $\omega_{pd}=200\,\mathrm{s}^{-1}$,
$\lambda_D\approx 45\,\mathrm{\mu m}$, and $C_d=9\,\mathrm{mm/s}$.
The observed nonlinear DA solitary pulses in Fig. 1(c) of Ref. \cite{Teng} had a periodicity of about $2\,\mathrm{mm}$, a mean speed of about $45\,\mathrm{mm/s}$, and a crest width (measured at the height of $N$ where $N=1$)
in the range $0.4$ -- $0.5,\mathrm{mm}$, with higher amplitude pulses having smaller widths.
We believe that there are some uncertainties in the plasma parameters that could explain the relatively low value of $C_d$ compared to the observed wave speed:
Increasing the values of $T_i$ to $0.10\,\mathrm{eV}$ and using the dust charging equation
[e.g. Eq.~(11) of Shukla and Eliasson \cite{Shukla09}], we obtain $Z_d=13800$ for $n_i=10^9\,\mathrm{cm}^{-3}$ and
$n_e=4.9\times10^8\,\mathrm{cm}^{-3}$, giving $\omega_{pd}=540\,\mathrm{s}^{-1}$, $\lambda_D=76\,\mathrm{\mu m}$, and
$C_d=41\,\mathrm{mm/s}$, which is compatible with the experiment. Using these parameters in our model, we have
$P=0.51$ and $\tau=0.025$, which we use in the simulation of the time-dependent system of equations (2)--(4).
The results are displayed in Fig. \ref{Fig:Teng}. We drive the DAW resonantly by an  external force of the
form $F=-0.01\sin[2\pi(X-T)/L]-0.001\sin[2\pi(X-T)/5L]$, added
to the terms in  the square parentheses in Eq. (3), where $L=26.3$ is the observed wave periodicity (2 mm) normalized
by $\lambda_D$. The result in Fig. \ref{Fig:Teng} shows almost periodic wave-trains
that develop into narrow peaks, very similar to the ones observed by Teng {\it et al.} \cite{Teng},
with density maxima about twice the ambient density and a typical width of about 4-5 Debye radii corresponding to
about 0.3-0.4 mm.
These spikes may be interpreted as driven large amplitude solitary DAW structures due to a balance
between the harmonic generation nonlinearities of the system and the dispersion provided by
the departure from the quasi-neutrality condition.

Bandyopadhyay {\it et al.} \cite{Band} studied how the speeds of DA solitary pulses depend on their amplitudes. The experimental plasma parameters were $n_i=7\times 10^{13}\,\mathrm{m}^{-3}$, $T_i=0.3\,\mathrm{eV}$, $T_e=8\,\mathrm{eV}$, $n_d=10^{10}\,\mathrm{m}^{-3}$, $Z_d=3\times 10^3$, $m_d=10^{-13}\,\mathrm{kg}$, giving $\omega_{pd}=51\,\mathrm{s}^{-1}$,  $\lambda_D=490\mathrm{\mu m}$ and $C_d=25\,\mathrm{mm/s}$, which corresponds to $P=0.43$ and $\tau=0.038$ in our model. The DA solitary pulses propagated with super-dust acoustic speeds, increasing with increasing amplitudes. In Fig. 1(b) of Ref. \cite{Band} a pulse of 100\% density amplitude propagates about $8\times10^{-3}\mathrm{m}$ in $0.24\,\mathrm{s}$,  giving a mean speed of $v_d\approx 0.033\,\mathrm{m/s}$, which corresponds to $M=v_d/C_d=1.33$. Figure \ref{Fig:Band} shows a simulation result, where the initial condition consists of a wide pulse of the  form $N=1+0.5\exp[-(X-20)^2/100]$, $U=0.5\exp[-(X-20)^2/100]$. The DA pulse breaks up into three DA solitary wave structures  propagating with the super-dust acoustic speed $M>C_a=1$.  Small but finite amplitude DA solitary pulses have the density profile $ N=1+N_0\, {\rm sech}^2(C_0^{1/2}\zeta/2)$, and the associated DAW potential
$\Phi=- (M^2 P N_0/\gamma){\rm sech}^2(C_0^{1/2}\zeta/2)$, where $N_0={3 C_0\gamma}/{2B M^2 P}$ is the amplitude, $C_0=1-1/M^2$, and $B=(1/2\gamma)[(1-P)\tau^2-1+3\gamma^2/M^4 P]$. Figure \ref{Fig:Band}(b)  exhibits  that the numerically obtained amplitudes of the three DA solitary pulses compare favorably well with the theoretical amplitude $N_0$. A pulse with a density amplitude of $N_0=2$ would have a speed of $M\approx 1.3$, which is also in good agreement with the experiment of Ref. \cite{Band}.

\section{Summary and conclusions}

In summary, we have presented a fully nonlinear, non-stationary unified theory for arbitrary large amplitudes DA shocks and DA solitary pulses  in a strongly  coupled dusty plasma. Our nonlinear theory is based on the Boltzmann distributed inertialess warm electrons and ions,  Poisson's equation, the dust continuity equation, and the generalized viscoelastic dust momentum equation for strongly correlated charged dust grains.
The governing nonlinear equations have been numerically solved to obtain the profiles of nonlinear DA waves, including the development of the DA shocks and DA solitary pulses. A comparison between our simulation results and  recent experimental observations \cite{Heinrich,Merlino} of the DA shocks in laboratory  dusty plasma discharges reveals a very good agreement with respect to the nonlinear DA wave speeds and DA shock wave smoothing  due to strong coupling effects between charged dust particles. From the width
of the DA shocks, one may, as suggested by Heinrich {\it et al.}~\cite{Heinrich}, infer the dust fluid viscosity. Furthermore, our simulation results of large  amplitude DA solitary pulses also compare favorably well with the observations of Bandyopadhyay {\it et al.} \cite{Band} and  Teng {\it et al.} \cite{Teng}.
Future experiments of nonlinear DAWs with higher precision measurements of the plasma parameters would be very valuable to benchmark the theoretical model. In closing, we stress that our fully nonlinear unified theory for DA shocks  and DA solitary pulses remain valid for a dusty plasma with a weak magnetic field (of the order of 100 Gauss), since the latter is unable to magnetize micron-sized charged dust particles and would not affect the trajectories of electrons and ions that follow the Boltzmann law on the spatio-temporal scales of our interest. A weak magnetic field just provides confinement for the electrons, which are coupled with  ions and negative dust grains through the space charge electric field of the DAW.

\acknowledgments
This research was supported by the Deutsche Forschungsgemeinschaft (DFG), Bonn, through the project SH21/3-2 of the  Research Unit 1048.

\end{document}